\documentstyle{l-aa} 
\input{psfig}
\begin{document} 
\thesaurus{11{(11.09.2; 11.11.1; 03.13.4)}}
\title{Determination of orbital parameters of 
       interacting galaxies using a genetic algorithm. \\
       {\large{Description of the method and application to artificial data.}}} 
\author{M. Wahde} 
\institute{NORDITA, Blegdamsvej 17, 2100 Copenhagen, Denmark.} 
\date{Received date; accepted date} 
\maketitle
\markboth{M. Wahde: Determination of orbital parameters of interacting galaxies} 
{M. Wahde: Determination of orbital parameters of interacting galaxies}
\begin{abstract}
A method for determining the orbital parameters of 
interacting pairs of galaxies is presented and evaluated using
artificial data. The method consists of a genetic algorithm which
can search efficiently through the very large space of possible orbits.
It is found that, in most cases, orbital parameters close to the
actual orbital parameters of the pair can be found. The method does not
require information about the velocity field of the interacting system,
and is able to cope with noisy data. The inner regions of the galaxies,
which are difficult to model, can be neglected, and the orbital parameters
can be determined using the remaining information.
\keywords{Galaxies: interactions - Galaxies: kinematics and dynamics -
Methods: numerical}
\end{abstract}
\section{Introduction}
Observations of interacting galaxies provide important information about
galactic structure and evolution. Due to the long time scales (compared to
the lifetime of a human) involved in interactions between galaxies, one can
only obtain a single snapshot of an interacting system. Dynamical modelling of
such systems therefore constitutes an important complement to observations, and can
naturally be divided into two parts: Modelling of the individual galaxies
participating in an interaction, and determination of the orbital parameters. 
Advances in the techniques for modelling of individual galaxies in interacting
systems has led to a greatly increased understanding of features such as bars, rings,
and spiral arms. 

In order to understand the dynamics of an interacting system of 
galaxies it is, however, equally important to know the parameters of the
relative orbit of the two galaxies. Determination of orbital parameters has only been
carried out for a small fraction of all observed interacting galaxies. 
An example of a system that has been much studied is the M51 system 
(for recent results, see Engstr\"om \& Athanassoula \cite{enat}, Howard \& Byrd 
\cite{hoby}, and Hernquist \cite{hern}). The fact that several authors have arrived at 
somewhat different orbits shows the difficulties involved in modelling even a 
well-observed system such as M51.

Some of the main problems encountered when numerical simulations are used 
for determining orbital parameters are that the parameter space that needs 
to be searched is often very large, and that the results of each simulation must 
be compared with the observational data. While methods for automatic comparison of
data between observations and simulations have been used in some cases (e.g.
Engstr\"om and Athanassoula \cite{enat}), very little has been done to find an efficient 
method for reducing the amount of searching necessary (i.e. the number of simulations) 
in order to find the orbit in a general case. In this paper, an efficient search method will 
be presented and evaluated using artificial data.

Science often benefits from the sharing of information between different
disciplines. One example of this is the invention of
{\em genetic algorithms} (Holland \cite{holl}). With natural evolution as the
inspiration, genetic algorithms (hereafter GAs) use artificial selection and
the genetic crossover and mutation operators to manipulate strings of numbers
which encode the variables of the problem, thereby reaching better and better
solutions to the problem. GAs are used in many branches of science (see the 
Appendix). However, in astrophysics there have, as yet, only been a few 
applications of GAs, in the fields of solar coronal modelling (Gibson \& 
Charbonneau \cite{gibs}), helioseismology (Tomczyk et al. \cite{tcst}), 
pulsar planet searching (Lazio \cite{lazi}), eclipsing binary stars (Hakala \cite{haka}), 
and gamma-ray astronomy (Lang \cite{lang}). For an excellent review of GAs in 
astronomy and astrophysics, see Charbonneau (\cite{char}). 

In this paper, a GA will be used for searching the space of possible orbital
parameters for pairs of interacting galaxies. Sect. \ref{descsec} contains a description
of the problem, and the method of solution is presented in Sect. \ref{methsec}. 
The results are given in Sect. \ref{ressec}, and are discussed in 
Sect. \ref{discsec}. The conclusions are presented in Sect. \ref{concsec}. 
The Appendix contains a brief description of the essential features of the 
GA used in the paper, as well as some references for further reading.
\section{Description of the problem}
\label{descsec}
The problem to be solved is the following: Given (photometric) observations 
of a pair of interacting galaxies, as well as systemic velocities for the
two galaxies, determine the parameters of the orbit.

In principle, three observations at different times suffice to deduce the 
orbital parameters of a comet or an asteroid. For galaxies, however, one
can only obtain observations of a single snapshot. Fortunately, such 
snapshots contain a wealth of information since the gravitational forces
between the galaxies produce deformations in the form of, for example, 
arms, tails, and bridges connecting the galaxies. In addition to the
position data, the radial velocity field can also be measured.  
From the complete set of data, information about scale radii, scale 
heights, disc inclinations, velocity dispersions, and masses can be obtained. 

However, positions along the line of sight and velocities in the plane
of the sky cannot be measured, and the problem of finding the orbital
parameters is therefore far from trivial. Additional complications
appear since observations never provide perfect, noise-free data. 

\begin{figure}
\centerline{\psfig{figure=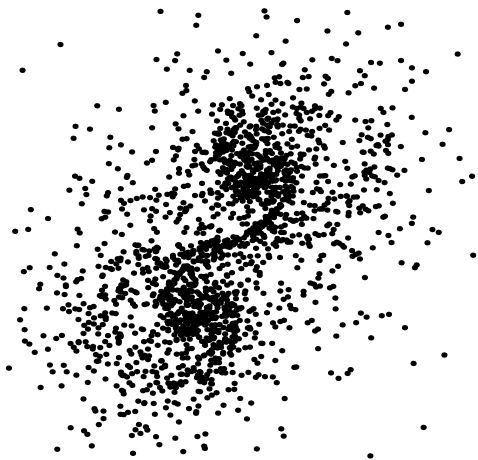,height=8.5cm}}
\caption{A snapshot of a pair of interacting galaxies. The corresponding
 orbital parameters are given in the bottom row of Table \ref{run1tab}.}
\label{obs1fig}
\end{figure}

A snapshot from a simulation of two face-on galaxies is shown in Fig. \ref{obs1fig}.
Clearly, the galaxies are strongly interacting with material being torn
of from both. Just looking at the snapshot, it is difficult to say 
very much about the orbit. Given systemic velocities and positional 
information of the type shown in Fig. \ref{obs1fig}, we shall 
in Sect. \ref{ressec} use a GA to determine 
the orbital parameters of several interacting systems.
\section{The method}
\label{methsec}
In this section, the method for determining orbital parameters will be 
presented. A general description of GAs is given in the Appendix, which also
contains references to more complete discussions of GAs.
\subsection{The simulations}
\label{methsim}
We have used a coordinate system such that the $x-y$-plane is the plane of the sky  
(with the $x-$axis horizontal), and the $z$-axis pointing {\em towards} the observer.
The longitude of the ascending node ($\Omega$) is measured from the $x-$axis.
The orbital inclination is defined to be zero if the orbit lies in the plane of the sky.

As unknown variables we have taken the relative (systemic) velocities between the galaxies in
the plane of the sky ($\Delta v_{x}$ and $\Delta v_{y}$), the separation along the
line of sight ($\Delta z$), the masses ($m_{1}$ and $m_{2}$), and the spins (clockwise
or counterclockwise) of the two galaxies. 
The encoding of the unknown parameters in the strings used by the GA is described 
in Sect. \ref{methgen}.

Observations provide values of the relative systemic velocities along the line of 
sight ($\Delta v_{z}$) and the separations in the plane of the sky 
($\Delta x$ and $\Delta y$).  

For each simulation, values of the unknown variables are obtained as described
in Sect. \ref{methgen} below, and the orbital parameters can then be determined from 
the masses ($m_{1}$ and $m_{2}$), the three 
components of the separation vector ($\Delta x$, $\Delta y$, and $\Delta z$), and 
the three components of the relative 
velocity ($\Delta v_{x}$, $\Delta v_{y}$, and $\Delta v_{z}$).
The orbital parameters are $a$ (semi-major axis), $e$ (eccentricity), $i$ (inclination), 
$\Omega$ (longitude of the ascending node), $\omega$ (the argument of pericentron),
and $T$ (the time of the pericentre passage). The transformations from 
cartesian coordinates and velocities to the orbital parameters is straightforward and well 
known, and will not be given here. For a discussion of the transformations involved,
see e.g. Danby (\cite{danb}) or Deutsch (\cite{deut}).

The orbital parameters are of course not absolutely necessary for orbit integration
- the cartesian coordinates and velocities would suffice - but the
orbital parameters have the advantage of giving information which is easier to
visualize. For instance, the value of $e$ immediately shows whether an 
orbit is elliptical or hyperbolic. 

When the orbital parameters have been found, the positions of the two galaxies are 
integrated backwards in time. During the backwards integration, the galaxies are
represented by point particles moving on a two-body orbit. Starting at
time $t_{0}$ the backward integration lasts for $N_{\rm step}$ steps. At each
step $j$, the new time is computed as $t = t_{0} - j {\rm d}t$, where ${\rm d}t$ 
is the length of the time step, and Kepler's equation is then solved.
The form of Kepler's equation is
\begin{equation}
E - e\sin E = \sqrt{\frac{G(m_{1}+m_{2})}{a^3}}\left( t-T \right),
\end{equation}
(where $E$ is the eccentric anomaly) for elliptical orbits and
\begin{equation}
e\sinh F - F = \sqrt{\frac{G(m_{1}+m_{2})}{(-a)^3}}\left( t-T \right),
\end{equation}
(where $F$ can be called the hyperbolic anomaly) for hyperbolic orbits. The case 
$e=1$ (exactly), i.e. parabolic orbits, is neglected. However, $e$ can be arbitrarily
close to 1.   

We use sign conventions such that $a < 0$ for hyperbolic orbits.
Using the orbital parameters, the relative positions and velocities
of the two galaxies can be obtained, through the transformations discussed in e.g.
Danby (\cite{danb}) or Deutsch (\cite{deut}).

At the end of the backward integration, a disc of particles is added to each of the
galaxies. Before the first simulation is carried out, the program reads two copies of a 
standard disc of unit mass and unit scale length. However, the masses and scale lengths 
of the two galaxies are generally different from unity, and therefore the positions and 
velocities of the particles in each disc are, for each simulation, scaled to appropriate values.

Then, the orbit is integrated forward in time until the final step (corresponding
to the time of the observation) is reached, at which point the position data are 
stored in the manner described below, and the next simulation can begin.

For hyperbolic orbits the backward integration is straightforward and can be
terminated when the galaxies are at sufficient distance (a few galactic radii, say) 
from each other. The situation for elliptical orbits is more difficult: If the
duration of the backward integration is badly chosen, the galaxies may not be
sufficiently separated at the start of the forward integration. To avoid this
problem, the value of $N_{\rm step}$ can be chosen individually for each orbit in
such a way that the backward integration proceeds until the apocentre is reached.
However, the problems encountered for elliptical orbits are more general than that,
since previous encounters may have damaged the discs of the galaxies, especially
for short period orbits. For this reason, we shall restrict ourselves to considering 
artificial data corresponding to hyperbolic orbits only. 
\subsection{Evaluation of the simulations}
\label{metheval}
When the final step has been reached, the output from the simulation should be compared
with the (artificial, in this paper) observational data. The data from an observation
can be either in the form of a contour map, or in the form of a grid of grayscale pixels,
the shading at each pixel determined by the amount of light at that point. The version
of the method described in this paper requires the data to be of the latter form, 
even though the program could be generalized to operate on contour maps as well.

Since we only use artificial observations here, we will use as data the amount of mass 
at each pixel rather than the amount of light. When real data is used, a scale factor (the
mass-to-light ratio) must be introduced. 

Thus, at the end of each run a grid is superposed on the two galaxies, and the amount
of mass in each grid cell is stored, each particle contributing $m_{1}/N_{p, 1}$ if it
belongs to the first galaxy or $m_{2}/N_{p, 2}$ if it belongs to the second,  
where $N_{p,1}$ and $N_{p,2}$ are the number of particles used for the first and the
second galaxy, respectively. 

The data from the orbit integration which provides the (artificial) observation 
is stored in the same way, and is read by the program before the first simulation 
is carried out. 

Note that the GA {\em does not} make use of the radial velocity field of the
two interacting galaxies: The only velocity information used are the systemic 
velocities of the two galaxies. The radial velocity field could be included in
the evaluation procedure, but that would put an unnecessary limit on the number 
of observed systems for which modelling could be carried out, since accurate velocity 
data are difficult to obtain.
\begin{figure*}
\centerline{\vbox{\psfig{figure=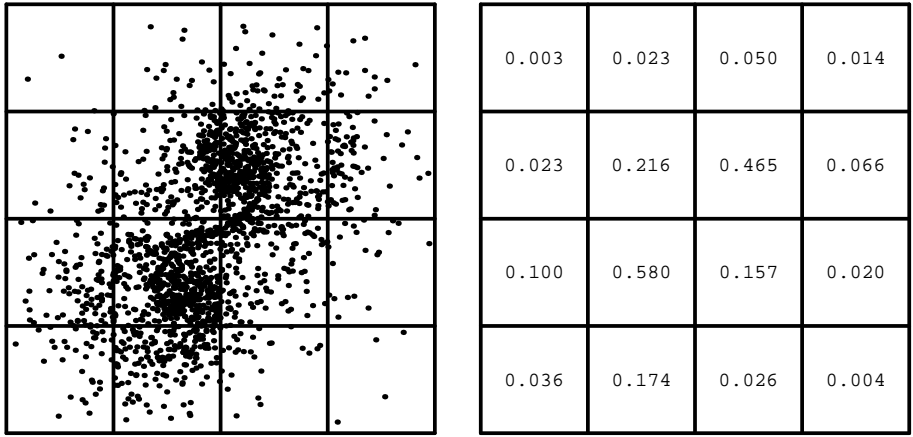,height=8.5cm}}}
\caption{Computation of the data for the observation in Fig. \ref{obs1fig}. 
The grid is superposed
on the image of the interacting galaxies, and the grid cells are assigned values 
corresponding to the mass in each cell. For clarity, only a very coarse 
grid has been used in this figure.}
\label{gridfig1}
\end{figure*}

The number of grid points in the $x-$ and $y-$directions, denoted $n_{x}$ and $n_{y}$,
as well as the size, $L \times L$, of the (quadratic) grid cells are input parameters
to the program, and should obviously be chosen in such a way that the relevant parts of the
two galaxies are contained within the grid. For the data presented in Fig. \ref{obs1fig}, 
the $x-$ and $y-$coordinates ranged from -14.0 to 14.0 in program units, so if the grid 
parameters are taken to be, for example, $n_{x} = n_{y} = 4$ and $L = 7.0$, the 
corresponding data will be matrix of mass values in the right of Fig. \ref{gridfig1}.  

In order to evaluate a given simulation, the deviation between its results and the
observational data should be measured. The deviation measure can be defined in 
different ways, and it will here be defined as 
\begin{equation}
\delta = \frac{n_{\rm typ}^2}{n_{x} n_{y}}
\sum_{i,j}|m_{i,j} - m^{\rm obs}_{i,j}|/(m_{\epsilon} + m^{\rm obs}_{i,j}),
\end{equation}
where $m_{i,j}$ is the mass in cell $(i,j)$ obtained from the simulation, 
$m^{\rm obs}_{i,j}$ is the same quantity obtained from the observational data,
and the sum extends over the whole grid. The $m_{\epsilon}$ in the denominator is
needed to prevent a divergence in the cases where $m^{\rm obs}_{i,j}$ is 
zero, and its value is taken to be the typical mass of a particle in a galaxy
of unit mass, i.e. $1/N_{p}$ where $N_{p}$ is the number of particles used in
the simulations. The factor in front of the summation sign is a normalization 
factor. Its sole purpose is to make $\delta$ independent of the number of grid points
used in the data comparison. Thus, $n_{\rm typ}^2$ is a typical value of the
number of pixels, here taken to be 49, and $n_{x} n_{y}$ is the number of pixels
used in the computer run. In the runs discussed in Sects. \ref{ressec} 
and \ref{discsec}, $n_{x} = n_{y} = 7$, yielding a normalization factor equal to 1. 
This choice is somewhat arbitrary. However, the values of $n_{x}$ and $n_{y}$ must 
neither be too small nor too large since, in the former case, the tidal features used 
by the GA will not be resolved and, in the latter case, the data from the observation 
will be unnecessarily detailed and thereby more sensitive to the noise which is always
present in real data.

When real data is used, it is the amount of light at each grid point, 
rather than the mass, that is detected, and the deviation measure could instead be 
defined as, for example,
\begin{equation}
\delta \propto \sum_{i,j}|g_{i,j} - g^{\rm obs}_{i,j}|/(\nu + g^{\rm obs}_{i,j}),
\end{equation}
where $g$ denotes the shading of pixel $(i,j)$. If the $g$ values were
normalized to lie between 0 (black, no light) and 1 (white, maximum light), the 
value of $\nu$ could be, say, 0.001. The deviation measures just defined punish strongly 
those simulations which try to put many particles in regions which are only supposed
to contain a few. 
\subsection{The genetic algorithm}
\label{methgen}
\begin{figure*}
\centerline{\psfig{figure=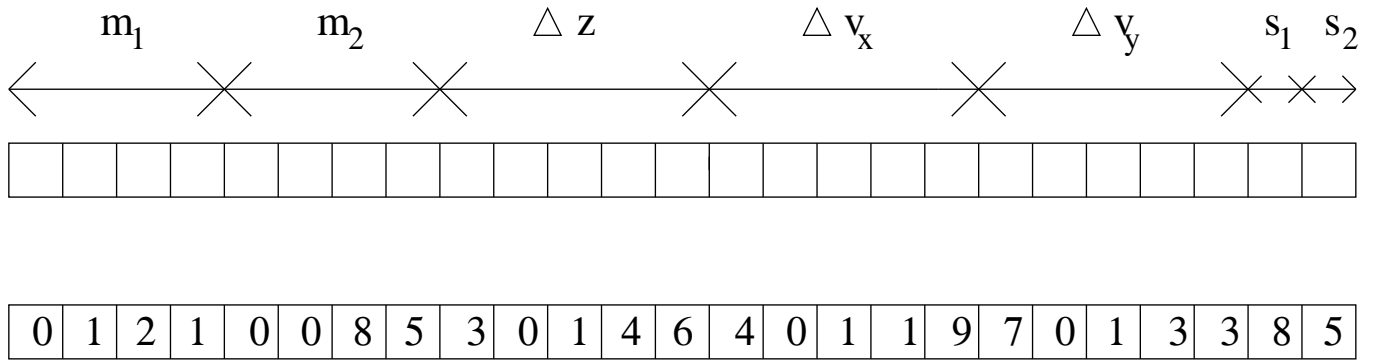,height=6.0cm}}
\caption{Top: The encoding scheme for the seven unknown variables. Genes 1-4 encode the
mass of the first galaxy such that $m_{1} = g_{1}*10^{1} + g_{2}*10^{0} + g_{3}*10^{-1}
+ g_{4}*10^{-2}$, where $g_{i}$ denotes the value of gene $i$. The other variables are 
encoded in a similar fashion. Genes 9, 14, and 19 encode the signs of $\Delta z$, 
$\Delta v_{x}$, and $\Delta v_{y}$, respectively,
such that the sign is negative if the corresponding value is odd and positive if it is even. The
spin of the first galaxy is $-1$ (clockwise) if the value of gene 24 is
odd, and $+1$ if it is even. The spin of the second galaxy is encoded, in the
same way, in gene 25. Bottom: an example of a chromosome. Decoding the chromosome, the values 
$m_{1} = 1.21$, $m_{2} = 0.85$ etc. are obtained.}
\label{gencodefig}
\end{figure*}

In the preceding subsections, the individual simulations were described. In this
subsection we shall see how the GA operates. At the start of each GA run, the
chromosomes of the $N_{\rm pop}$ simulations (or {\em individuals} in the biological
terminology used in the Appendix) of the first generation are initialized by 
assigning random numbers between 0 and 9 to each of the genes. The encoding 
of the variables is illustrated in Fig. \ref{gencodefig}. As an example of the decoding, the
string at the bottom in Fig. \ref{gencodefig} would, when decoded, give 
the values $m_{1} = 1.21$, $m_{2} = 0.85$, $\Delta z = -14.6$, 
$\Delta v_{x} = 0.119$, $\Delta v_{y} =
-0.133$, $s_{1} = 1$, and $s_{2} = -1$. Note that the encoding is decimal
rather than binary. 

When the chromosomes have been initialized, the program loops through all of the
$N_{\rm pop}$ individuals in the first generation. For each individual, the
chromosome is decoded and the orbital parameters are computed as outlined in Sect.
\ref{methsim}. Then, the two orbit integrations (backward and forward) are carried out, 
the result is compared with the observational data in the manner 
described in Sect. \ref{metheval}, and the resulting value of the deviation is 
computed and stored. As a measure of the fitness of an individual we use the function
\begin{equation}
\label{fiteq}
f = \frac{1}{1+\delta},
\end{equation}
i.e. essentially the inverse of the deviation. $f = 1$ corresponds to a perfect match.
When all individuals of the first generation have been evaluated, the fitnesses are
ranked using linear fitness ranking as described in the Appendix. Thus the exact form
of the mapping from $\delta$ to $f$ will be of no importance when new generations are
formed, and any fitness function for which the fitness increases with decreasing $\delta$ 
could have been used. However, as discussed in Sect. \ref{resothers} below, the fitness values
obtained from Eq.(\ref{fiteq}) play an important role when the quality (i.e. the goodness
of fit) of a set of orbital parameters is evaluated.
 
In order to produce the genetic material, i.e. the chromosomes, of the second generation, 
the chromosome of the best individual in the first generation is copied twice, and the 
remaining $N_{\rm pop}-2$ individuals of the second generation are formed through the 
process of parent selection 
followed by crossover and finally mutation as described in the Appendix. When all $N_{\rm pop}$ 
new individuals have been generated, the genetic material of the first generation is deleted 
and the evaluation of the second generation can begin. 
After the second generation has been evaluated, the third generation is formed, all
its constituent individuals are evaluated, etc.

In this process, individuals with high fitness values  
will have a greater chance of spreading their genetic material to the next generation than 
individuals with low fitness values. To those unfamiliar with GAs, this process may not,
at a first glance, seem to be very efficient. However, as many authors have concluded
(e.g. Charbonneau \cite{char}) and as we shall see in Sect. \ref{ressec}, 
GAs are in fact {\em very} 
efficient for solving optimization problems for which the search spaces are large.
\subsection{Notes on the simulations}
\label{methnot}
The purpose of this paper is to test GAs as a method for finding orbital parameters of
interacting galaxies, rather than testing the methods used for the individual simulations.
Since many computer runs, some of which are described in Sects. \ref{ressec} 
and \ref{discsec},
had to be carried out during the testing procedure, some simplifications were introduced. 
It should be noted, however, that these simplifications can be removed so that, when the 
orbit of a single observed interacting pair is to be computed, more advanced simulation methods 
employing e.g. greater numbers of particles can be used. 

Thus, the simulations were non-self-gravitating, i.e. interparticle forces were neglected and 
the disc particles were influenced only by the gravitational forces from the two point particles
(of mass $m_{1}$ and $m_{2}$), representing each galactic disc. The scale length ($r_{d}$) of each 
disc was assumed to be known from measurements of the observational data, the discs were
assumed to be face-on, and no dispersion velocities were added. Since the galaxies were
face-on in all simulations, the scale height was of little importance and was set equal to 
$r_{d}/10$. In general, the discs consisted of 1,000 particles each for a total of 2,000
particles per simulation. The exception is Run 4 (see below) for which a total of 10,000 particles 
were used. The (exponential) scale lengths, (${\rm sech}^2$) scale heights, 
velocity dispersions, and orientations of the discs can all be added to the set of variables 
by lengthening the chromosome of each individual. As the number of variables increases, the 
search times will also increase. However, the increase in 
search time for a GA will be much smaller than the corresponding increase for 
the full parameter space searches discussed in Sect. \ref{disccomp} below, 
and the case for using a GA will then be even stronger.

The neglect of self-gravity may not be a serious limitation, in the cases where only the 
outer parts of the two galaxies are significantly affected by the interaction: In a 
tenuous, low mass arm or tail consisting of material from the outer regions of either 
galaxy, the self-gravity is usually less important than the tidal field of the two galaxies. 
In any case, non-self-gravitating simulations were sufficient for the purposes of this
paper.

A typical 100 generation computer run with 1,000 particles per galaxy and 
$N_{\rm pop} = 500$, requires 12.3 CPU hours on a Sun Enterprise 2.

\subsection{Units and parameter intervals}
\label{methunit}
We have used units such that $G$, the gravitational constant, is equal to 1. The unit
of length can be taken to be 1 kpc, the unit of time $1.05$ Myr, and the unit of 
mass $ 2\times10^{11}M_{\odot}$. The unit of velocity is then equal to 931 km/s. 
Other scalings to physical units can, of course, be used as well.
In the genetic encoding scheme, the galaxy masses range from 0.01 to 99.99, the separation
along the line of sight ($\Delta z$) ranges from -999.9 to 999.9, and the relative velocities
($\Delta v_{x}$ and $\Delta v_{y}$) range from -9.999 to 9.999. The spins can take the
values -1 and 1. The total number of combinations is $3.2 \times 10^{21}$~(!). 

Often, however, the full ranges just described need not be used. For instance, if
two galaxies are strongly interacting and connected by a bridge, one does not expect them
to be almost a thousand kpc apart along the line of sight! If, for instance, the 
values of $\Delta z$ can be restricted to the interval -50 to 50, say, the number of possible
values of $\Delta z$ is reduced from 20,000 to 1,000. Even with such reductions, the
search space will still be very large, and an efficient search method is therefore
needed.
\section{Results}
\label{ressec}
\subsection{Run 1}
\label{resrun1}
\begin{figure*}
\centerline{\psfig{figure=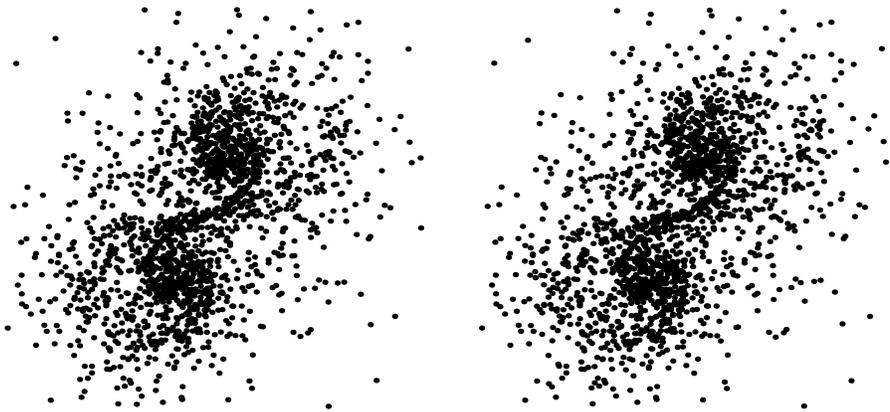,height=8.5cm}}
\caption{The best simulation in Run 1 (left panel) and the observation (right panel).}
\label{compfig}
\end{figure*}

In this section, the results of applying the method presented above will be given, 
starting with the system shown in Fig. \ref{obs1fig}. In order to determine the 
corresponding orbital parameters, a run with population size $N_{\rm pop} = 500$
was carried out. The number of generations ($N_{\rm gen}$) was 100, the mutation rate was
$p_{\rm mut} = 0.003$, and the number of grid cells was 49 ($n_{x} = 7, n_{y} = 7$).
For Run 1 and all other runs described here, the values of $m_{1}$ and $m_{2}$ were 
constrained to lie between 0.3 and 3.0, $\Delta z$ between -50.0 and 50.0, 
and $\Delta v_{x}$ and $\Delta v_{y}$ between -0.999 and 0.999. All possible spin
combinations were allowed. The total number of combinations of the unknown variables 
$m_{1}, m_{2}, \Delta z, \Delta v_{x}, \Delta v_{y}, s_{1}, s_{2}$ was then 
approximately $1.2 \times 10^{15}$.
Mutated chromosomes were only accepted if their values of the unknown
parameters were contained in the above intervals. 
For the observational data used in Run 1, the actual values of $m_{1}$, $m_{2}$, 
$\Delta z$, $\Delta v_{x}$, $\Delta v_{y}$, $s_{1}$, and $s_{2}$ were
1.0, 1.0, 3.0, -0.672, 0.839, 1, and 1, respectively.
Table \ref{run1tab} presents the results of Run 1: The first 6 rows
show the orbital parameters of the best simulation in generations 1, 5, 10, 20, 50, and 100, 
and the final row shows the actual orbital parameters of the \lq observed\rq \, system. As can 
be seen, the GA was able to find the orbital parameters with great accuracy.
In fact, acceptable orbital parameters were obtained already after 20 generations. 
Fig. \ref{compfig} shows the final configuration of the best individual in 
generation 100 (left panel) together with the data from the artificial observation
(right panel).
\begin{table*}[ht]
\caption[]{Results from Run 1. The first 6 rows show, for the best simulation of generations
1, 5, 10, 20, 50, and 100, the orbital parameters $a$, $e$, $i$, $\omega$, $\Omega$, and 
$F_{0}$, as well as the spins, $s_{1}$ and $s_{2}$, and the masses, $m_{1}$ and $m_{2}$, of 
the galaxies. 
The parameter $F_{0}$ is the hyperbolic anomaly at the final time step of each simulation, 
from which $T$, the time of the pericentre passage, can be 
computed. The final row shows the actual orbital parameters used for generating Fig. \ref{obs1fig}.}
\label{run1tab}
\begin{flushleft}
\begin{tabular}{lllllllllll}
\hline\noalign{\smallskip}
Gen. & $a$ & $e$ & $i$ & $\omega$ & $\Omega$ & $F_{0}$ & $s_{1}$ & $s_{2}$ & $m_{1}$ & $m_{2}$ \\
\noalign{\smallskip}
\hline\noalign{\smallskip}
1 & -1.247 & 11.34 & 52.85 & 293.9 & 121.9 & -1.239 & -1 & 1 & 0.91 & 1.05 \\
5 & -8.722 & 1.836 & 37.99 & 324.4 & 61.32 & 24.24 & 1 & 1 & 0.79 & 0.92 \\
10 & -12.49 & 1.687 & 45.65 & 314.7 & 81.96 & 12.09 & 1 & 1 & 0.90 & 1.05 \\
20 & -2.105 & 4.796 & 24.13 & 17.70 & 18.52 & 27.77 & 1 & 1 & 0.98 & 1.00 \\
50 & -2.097 & 4.594 & 23.49 & 13.48 & 16.55 & 33.21 & 1 & 1 & 0.99 & 1.00 \\
100 & -2.231 & 4.416 & 23.96 & 13.01 & 18.03 & 32.03 & 1 & 1 & 0.99 & 1.00 \\
\hline\noalign{\smallskip}
Obs. & -2.184 & 4.466 & 23.96 & 14.33 & 15.75 & 33.00 & 1 & 1 & 1.00 & 1.00 \\
\noalign{\smallskip}
\hline
\end{tabular}
\end{flushleft}
\end{table*}
\subsection{Additional runs}
\label{resothers}
The results of three additional runs are shown in Table \ref{othertab}. The upper row in each
pair shows the orbital parameters of the best simulation in the final generation,
and the lower row shows the actual orbital parameters of the (artificial) observation.
The grids used for data comparison were again of size $7 \times 7$, and the population size 
was equal to 500 for all runs. 1,000 particles per galaxy were used
in Runs 2 and 3. In Run 4, 5,000 particles per galaxy were used. The artificial 
observation used in Run 4 was generated with the same orbital parameters as those used 
for generating the observational data of Run 1, to facilitate comparison between the two runs.
As is evident from the table, acceptable orbital parameters were 
found in all cases.
\begin{table*}
\caption[]{Results from Runs 2 to 4. For each pair of rows, the upper row shows
the orbital parameters of the best simulation in generation 100, and the lower
row shows the orbital parameters corresponding to the observational data. Upper pair:
Run 2, middle pair: Run 3, lower pair: Run 4.}
\label{othertab}
\begin{flushleft}
\begin{tabular}{lllllllllll}
\hline\noalign{\smallskip}
Gen. & $a$ & $e$ & $i$ & $\omega$ & $\Omega$ & $F_{0}$ & $s_{1}$ & $s_{2}$ & $m_
{1}$ & $m_{2}$ \\
\noalign{\smallskip}
\hline\noalign{\smallskip}
100 & -8.886 & 2.215 & 47.04 & 334.3 & 307.3 & 33.40 & 1 & 1 & 1.10 & 0.89 \\
\hline\noalign{\smallskip}
Obs. & -8.193 & 2.409 & 44.08 & 337.3 & 310.8 & 27.86 & 1 & 1 & 1.08 & 1.02 \\
\noalign{\smallskip}
\hline
\hline\noalign{\smallskip}
100 & -16.26 & 1.501 & 70.10 & 112.9 & 309.2 & 28.58 & 1 & 1 & 1.30 & 0.51 \\
\hline\noalign{\smallskip}
Obs. & -13.83 & 1.589 & 67.76 & 113.5 & 309.6 & 30.07 & 1 & 1 & 1.28 & 0.54 \\
\noalign{\smallskip}
\hline
\hline\noalign{\smallskip}
100 & -2.196 & 4.477 & 24.36 & 15.90 & 14.73 & 32.57 & 1 & 1 & 0.99 & 1.00 \\
\hline\noalign{\smallskip}
Obs. & -2.184 & 4.466 & 23.96 & 14.33 & 15.75 & 33.00 & 1 & 1 & 1.00 & 1.00 \\
\noalign{\smallskip}
\hline
\end{tabular}
\end{flushleft}
\end{table*}

Not all determinations of orbits proceed as smoothly as, for example,
Run 1. In order for the GA to be able to find the orbit, it is a requirement
that the galaxies show some signs of interactions, i.e. distortions of some
form. This was the case for Runs 1 to 4. However, in a case where the observation consisted
of two more or less unperturbed discs, the GA was not able to find the orbit.
The fact that signs of interactions are needed is rather obvious and does not,
in practice, imply any restrictions. After all, the method is intended for
{\em interacting} systems.

A more important problem is that, even for clearly interacting systems, the GA is
not always able to find the correct orbit on the first attempt. Since the
population size is far from infinite, there may simply never be sufficient 
variation in the genetic material to obtain the correct orbital parameters,
at least if $p_{\rm mut}$ is small. $p_{\rm mut}$ can of course be increased,
but the larger its value, the more the GA approaches a random search. Fortunately,
it is always easy to distinguish between a run that fails and one that succeeds,
namely through the fitness values. The fitness measure defined in Eq.(\ref{fiteq})
has the property that, if $f \geq 0.15$, the orbital parameters match well those of the
observed system\footnote{Thus, the fitness increases steeply when the
final details of the fitting are carried out, and fitness values above 0.5 are
rarely reached.}. For the excellent fit obtained for Run 1, the fitness of the
best simulation was 0.430. The corresponding numbers for Runs 2 to 4 were
0.192, 0.339, and 0.475. 

In contrast, in a run that fails, fitness values above 
0.1 are not reached, and usually the GA gets stuck at even lower values. Even
in such situations improvements do occur, but at a very slow rate, and it
is usually faster to restart the run, using different values of the random
number generator seed and the mutation rate.

The results for Runs 1,2, and 4 were obtained on the first attempt, but Run 3
required two attempts. In the first attempt, with 
mutation rate $p_{\rm mut} = 0.003$, the GA got stuck at a suboptimal solution 
with fitness 0.0863. The mutation rate was then increased to 0.010 and the 
random number generator seed was changed, resulting in a maximum fitness of 0.339 in
the second attempt.
\subsection{Blocking out the inner regions of the galaxies}
\label{resblack}
\begin{figure}
\centerline{\psfig{figure=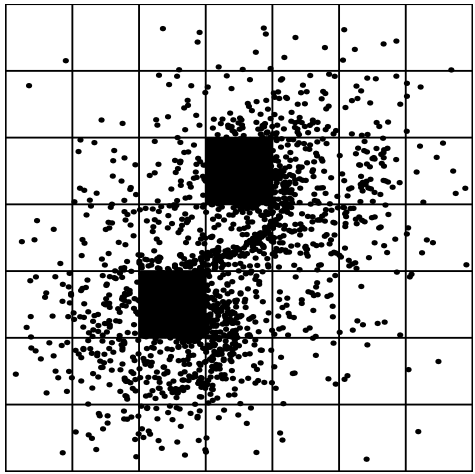,height=8.5cm}}
\caption{The two regions of highest density were discarded, as indicated by the
black squares. The original picture, before discarding the data in the two black squares, was
identical to Fig. \ref{obs1fig}.}
\label{blackfig}
\end{figure}

As mentioned previously, tidal features such as bridges and tails are used by the
GA when it attempts to determine the orbital parameters of an observed system. 
However, unlike the artificial data used here, real observational data is made
complicated by the existence of features such as bars, rings, ovals, etc. in
the inner regions of the galaxies. 
Furthermore, in an observation for which the tidal
features are clearly seen, the inner regions may be saturated, i.e.
overexposed. Thus, in order to avoid problems caused by the appearance of the inner
regions, a modified version of the GA program was constructed such that any pixel in
the grid could be blocked out and discarded. An example of a run with the modified program
is illustrated in Fig. \ref{blackfig}, where the two pixels with highest density were
discarded. 
\begin{table*}[ht]
\caption[]{Results from the run illustrated in Fig. \ref{blackfig}. The upper row shows
the orbital parameters of the best simulation in generation 100, and the lower
row shows the orbital parameters corresponding to the observational data.}
\label{blacktab}
\begin{flushleft}
\begin{tabular}{lllllllllll}
\hline\noalign{\smallskip}
Gen. & $a$ & $e$ & $i$ & $\omega$ & $\Omega$ & $F_{0}$ & $s_{1}$ & $s_{2}$ & $m_
{1}$ & $m_{2}$ \\
\noalign{\smallskip}
\hline\noalign{\smallskip}
100 & -2.483 & 4.126 & 25.99 & 17.78 & 13.37 & 31.69 & 1 & 1 & 0.99 & 1.00 \\
\hline\noalign{\smallskip}
Obs. & -2.184 & 4.466 & 23.96 & 14.33 & 15.75 & 33.00 & 1 & 1 & 1.00 & 1.00 \\
\noalign{\smallskip}
\hline
\end{tabular}
\end{flushleft}
\end{table*}

Thus, in this run, the deviation was computed using only the 47  
(i.e. $n_{x}\times n_{y} - 2$) remaining pixels. Clearly, this problem is more difficult to
solve for the GA (or anyone else!), since only partial information is accessible. 
The result of the run is shown in Table \ref{blacktab}, the upper row as 
usual showing the orbital
parameters of the best simulation and the lower row showing the observational data, which
were the same as for Run 1. Even though the resulting fit is not as stunning as for Run 1,
acceptable orbital parameters were obtained.
\section{Discussion}
\label{discsec}
\subsection{Comparison with other methods}
\label{disccomp}
As mentioned in the introduction, very little has been done to find
efficient search methods for the problem of determining orbital parameters in a 
general case. An exception is the work by Salo and Laurikainen (\cite{sala}) in which
an error minimization technique was used to fit the orbital parameters of
NGC 7753/7752. However, in their paper, only three variables were taken as fitting 
parameters for the error minimization technique and other variables, e.g. the masses, 
were kept fixed.

A common and straightforward method for finding orbital parameters is to carry
out a full search of (part of) the parameter space. 
The drawback with such a method is that the number of
simulations needed to achieve acceptable resolution over the parameter space 
grows very fast in an $N$-dimensional space where $N \geq 5$. A simple example
should suffice as illustration: In our Run 1, a very close match was obtained
after 100 generations consisting of 500 simulations each, i.e. after 50,000 simulations.
A full parameter search would only allow $(50000/4)^{1/5} \approx 6.59 
\approx 7$ values of each of the parameters $m_{1}, m_{2}, \Delta z, \Delta v_{x},$
and $\Delta v_{y}$. The factor $4$ in the denominator comes from the fact that,
for each set ${m_{1},m_{2},\Delta z, \Delta v_{x}, \Delta v_{y}}$, four values of
the spin must be tested. Even if the spins are assumed to be known, only 
$50000^{1/5} \approx 8.7 \approx 9$ values of each parameter could be tested. 
With the range $[-50.0,50.0]$ for $\Delta z$, as used in Run 1, that would give 
a resolution of only $100/(9-1) = 12.5$ and an equally poor resolution for the
the other parameters. Note also that, for Run 1, the GA found an acceptable solution already
after 20 generations, or 10,000 simulations. With 10,000 simulations available,
a full search would allow only $6.3 \approx 6$ values of each parameter to be
tested, if the spins were known.

Clearly, a complete, unbiased search of the whole parameter space is not an 
option, at least when the search space is of dimension five or higher. Thus, in order
to carry out a full search, either the number of dimensions must be reduced by fixing
the values of some parameters, or, if all parameter are used, some constraints must
be imposed on the values to be searched. While this is certainly possible in some 
cases, the constraints may not always be justified and it may be impossible to find the
correct orbit after imposing the constraints. A strong advantage of the GA method is 
that only very loose constraints need to be imposed, even when the 
number of parameters is large.

\subsection{Comparison with random search}
\label{discrand}
\begin{figure}
\centerline{\psfig{figure=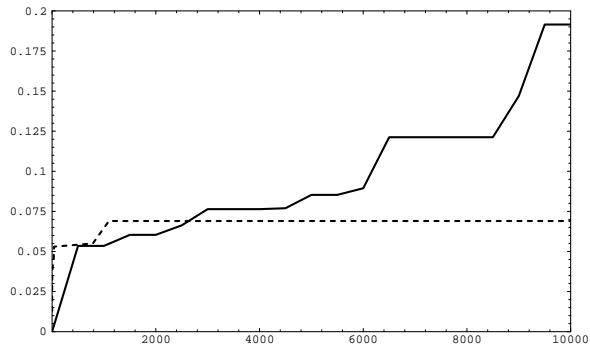,height=7.5cm}}
\caption{Comparison of the performance of a GA and a random search. The dashed
line corresponds to the random search. The vertical axis shows fitness values and the
horizontal axis the number of individuals evaluated.}
\label{rancompfig}
\end{figure}

Even though mutations (which play a subordinate role) are random, {\em selection} is not, 
and a GA is strongly different from a random search.
In order to illustrate the non-randomness of the GA method,
a calculation was carried out in which the values of the parameters 
$m_{1}, m_{2}, \Delta z, \Delta v_{x}, \Delta v_{y}, s_{1}$, and $s_{2}$ were generated 
at random. A total of 10,000 random sets of parameters were generated, corresponding to 
20 generations of 500 simulations for the GA. In Fig. \ref{rancompfig}, the results 
of the first 20 generations of Run 1 are compared with the results of the random run. 
The GA took the lead already after 6 generations, and the difference in performance 
should be obvious from the figure. 
\subsection{Sensitivity to noise}
\label{discnoise}
Whereas the artificial data sets used so far were noise-free, real data sets 
are invariably noisy. Furthermore, even if the noise levels are low, in real data there 
will always be other deviations from the idealized situation used in the GA simulations. 
For example, the mass-to-light ratio need not be constant.
\begin{table*}
\caption[]{Results from the two runs with noisy input data. The upper row shows
the orbital parameters of the best simulation in generation 100 for the run with
10 \% noise added, the middle row shows the same parameters for the run with
30 \% noise added, and the bottom row shows the orbital parameters corresponding to 
the observational data. Note that an exact match is impossible in this case, because 
of the noise added to the data.}
\label{noistab}
\begin{flushleft}
\begin{tabular}{llllllllllll}
\hline\noalign{\smallskip}
Gen. & $a$ & $e$ & $i$ & $\omega$ & $\Omega$ & $F_{0}$ & $s_{1}$ & $s_{2}$ & $m_
{1}$ & $m_{2}$ & Remark \\
\noalign{\smallskip}
\hline\noalign{\smallskip}
100 & -2.221 & 4.398 & 23.77  & 12.65 & 17.43  & 32.88 & 1 & 1 & 0.99 & 1.02 & $10 \%$ noise \\
\hline\noalign{\smallskip}
100 & -1.901 & 5.045 & 23.33 & 16.14 & 16.01 & 31.79 & 1 & 1 & 0.98 & 0.95 & $30 \%$ noise \\
\hline\noalign{\smallskip}
Obs. & -2.184 & 4.466 & 23.96 & 14.33 & 15.75 & 33.00 & 1 & 1 & 1.00 & 1.00 & \\
\noalign{\smallskip}
\hline
\end{tabular}
\end{flushleft}
\end{table*}
Therefore, in order to be useful, the GA method must be able to function even in
the presence of noise. 
We have tested the noise sensitivity by first adding $10 \%$ noise to the
data set used in Run 1. 
The noise was added by changing the values of the masses in the grid of
observational data according to
\begin{equation}
m_{i,j} \rightarrow m_{i,j}(1 + \alpha_{i,j}),
\end{equation}
where $m_{i,j}$ denotes the mass in cell $(i,j)$, and $\alpha_{i,j}$ are random
numbers between $-\alpha_{\rm max}$ and $\alpha_{\rm max}$, where $\alpha_{\rm max}$ is the
noise level. Thus, for the run with $10 \%$ noise, $\alpha_{\rm max}$ was equal to 0.1.
The results are shown in the upper row of Table \ref{noistab}, from which it
evident that 10 \% noise does not stop the GA from finding the correct solution.
The results from a run with $30 \%$ noise are shown in the middle row of
Table \ref{noistab}. Even in this case, the GA manages to find a solution fairly close
to the correct one.
\section{Conclusions and directions for further work}
\label{concsec}
We have presented an efficient method, based on genetic algorithms, for finding the
orbital parameters of interacting galaxies, and applied it to simulated galaxies on 
hyperbolic orbits. 
Given the centre-of-mass positions of both galaxies, their radial velocities, the scale radii 
of their discs, and a matrix of observed (light or mass) density data, the GA was able to find 
the orbital parameters with great accuracy in most cases. In the cases where an 
orbit could not be found, the failure could be detected rather quickly from the fitness values.

\begin{figure}[ht]
\centerline{\psfig{figure=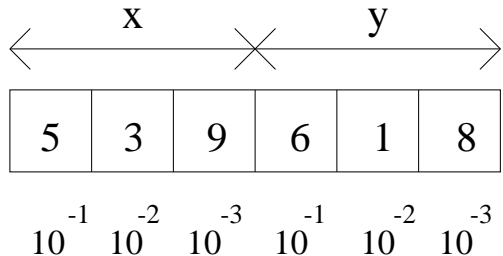,height=5.0cm}}
\caption{A typical chromosome of an individual in the inverse-function example. The first
gene codes for the first decimal of $x$, the second gene for the second decimal of $x$ etc.
Decoding the chromosome, the values $x = 0.539$ and $y = 0.618$ are obtained.}
\label{genomefig}
\end{figure}
The GA can operate on reduced data sets, in which the inner regions have been 
blocked out, enhancing the tidal features which are needed for the determination
of the orbital parameters. The method does not make use of the velocity field
of the interacting system. 

Even though there have been several simplifications in the test cases used here,
the GA method is a promising approach to the type of problems considered in the
paper. In order to make the method even more useful, the possibility of using
inclined discs, adding another 4 parameters to the set of unknowns, should also 
be considered. The scale radii of the two discs could also be included in the 
parameter set. Another improvement could come from using a simulation code 
incorporating the self-gravity of the two galaxies. Dark matter haloes
could also be added, and if the distributions of dark matter were chosen in
such a way as to be easily parametrized by one or a few parameters, it
would be possible to add also these parameters to the set of unknowns.

Still, even in its more primitive present state, the GA method can be very helpful
for finding orbital parameters. The output orbital parameters of the best
simulation in a GA run can be used as input parameters for an advanced 
self-gravitating $N$-body simulation incorporating gas dynamics and dark matter.
\section*{Appendix}
The aim of this Appendix is only to introduce and describe some of the most elementary 
features of GAs. For a much more complete discussion of GAs and their performance
compared with other algorithms, see Holland (\cite{holl}), Davis (\cite{davi}), 
or Mitchell (\cite{mitc}). 
For a review of GAs in astronomy and astrophysics, the article by Charbonneau (\cite{char}) 
is strongly recommended. 

GAs have been applied in many different subjects, including machine learning, population
genetics, neural network design, economics etc. Among other things, GAs are well suited
for search and optimization, and are particularly useful when the search spaces are large.

Since GAs are inspired by natural evolution, the terminology often involves terms from
biology, such as genes, populations, fitness etc. For an introduction to the terminology,
see e.g. Charbonneau (\cite{char}) or Mitchell (\cite{mitc}). 
Whenever such terms are introduced for the 
first time in this Appendix, they will be put in {\em italics}, and hopefully their 
meaning should be clear from the context.

When a GA is to be applied to an optimization problem, 
the variables of the problem are first encoded in strings (of given length) of integers. 
Initially, a {\em population} (i.e. a set) of $N_{\rm pop}$ {\em individuals} are formed 
by randomly generating such a string for each individual. 
Each string constitutes the {\em chromosome} (i.e. the genetic material) 
for the individual. 
The encoding can be either binary or decimal such that the values at
the different {\em genes} (i.e. locations) along the string 
are integers in the range [0,1] (for the binary case) or [0,9] (for the decimal case). 
The whole set of $N_{\rm pop}$ individuals with their corresponding chromosomes constitutes
the first {\em generation}. 

\begin{figure*}[ht]
\centerline{\psfig{figure=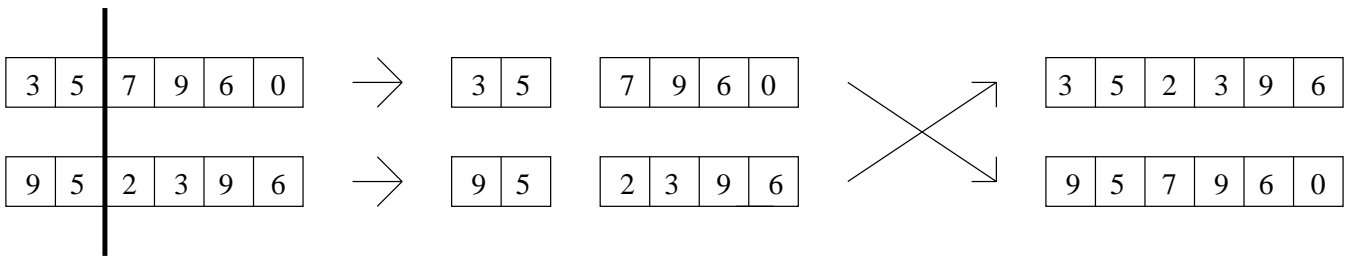,height=5.0cm}}
\caption{The crossover procedure. The chromosomes are divided at the crossover point, which
is indicated by a thick line, and the parts are joined as shown above.}
\label{crossfig}
\end{figure*}
As a trivial example, (the \lq inverse function example\rq), imagine that one wishes to find a 
pair of numbers $(x_{0},y_{0})$ such that a given function $h(x,y)$ takes a particular 
value $h_{0} = h(x_{0},y_{0})$. For simplicity, assume that $h(x,y)$ takes the 
value $h_{0}$ only at the one point $(x_{0},y_{0})$ and that $x$ and $y$ both lie in 
the interval [0,1]. If decimal encoding with three digit accuracy is used, the chromosome 
of an individual could have the form shown in Fig. \ref{genomefig}. 

When the first generation has been formed, the {\em fitness} of its constituent individuals
should be evaluated. Thus, for each individual, the variables are obtained by decoding the
chromosome. Given those variables, the relevant computation can be carried out. In the inverse
function example, the computation consists of forming $h(x,y)$ using the values of
$x$ and $y$. Then the result of the computation is compared with the desired result,
and a fitness value is assigned such that the smaller the deviation from the desired
result, the higher the fitness. 

If $h(x,y) = {\rm e}^{x y}$ in the inverse function example, and one is looking for
values $x$ and $y$ in [0,1] such that $h(x,y) = {\rm e}$ (the correct 
solution of course being $x=1$, $y=1$), then the individual shown in Fig.
\ref{genomefig} would give the value $h(0.539,0.618) = 1.39529$, the deviation 
would be $\delta = e - 1.39529$, and the corresponding fitness value $f$ could 
be defined as $f = 1/(1+\delta)$. 

When all the individuals of the first generation have been evaluated and fitness
values have been assigned, the second generation is formed by applying various
procedures inspired by natural evolution to the chromosomes of the individuals in 
the first generation.
These procedures include {\em selection} (followed by {\em crossover}) and {\em mutation}.

In order to perform a crossover between two chromosomes, two {\em parents} are selected
from the generation just evaluated. The choice of parents is made in such a way that
individuals with higher fitness have a greater probability of being selected than
individuals with lower fitness. The fitness values can either be used directly, or 
some more sophisticated method can be employed. Linear fitness ranking is one example
of such a method, in which the individuals are sorted according to their fitness and 
the best individual is assigned a new fitness equal to $N_{\rm pop}$, the second best 
is assigned a new fitness equal to $N_{\rm pop} - 1 $, and so on. This procedure enhances 
the differences between the individuals, especially if their original fitness values 
(before ranking) are very similar to each other. 

There exist several methods of choosing parents, and
here only one of the simplest shall be discussed, namely roulette-wheel selection. When
this selection method is used, the sum of the fitnesses $f_{i}, i = 1,2,...,N_{\rm pop}$ 
is formed, a random number $r$ between 0 and $\sum_{i}f_{i}$ is generated,
and the first individual $i$ which satisfies the condition 
\begin{equation}
\sum_{j=1}^{i}f_{j} \geq r,
\end{equation}
is selected as a parent. As an example, if $N_{\rm pop} = 3$ and the 
fitness values are 2, 5, and 3, the first individual is selected if
$r \leq 2$, the second is selected if $2 < r \leq 7$ and the third is selected if $r > 7$.
When two parents have been chosen (usually with replacement, i.e. such that an individual
can be chosen several times), crossover is performed by dividing the chromosomes of the
two parents into two parts, and joining the parts as shown in Fig. \ref{crossfig}. The point 
at which the cut is performed is called the {\em crossover point}.

When crossover is carried out, two partial solutions to the problem can be joined to
form a full solution. Returning to the inverse function example with
$h(x,y) = {\rm e}^{x y}$ as above, it is clear that the two parents in Fig. \ref{crossfig}
would both be rather far from the correct solution $x=1, y=1$. However, the second 
of the two new individuals (bottom row in the figure) formed by crossover would be much 
closer to the solution and would obtain a high fitness value.  
\begin{figure*}[ht]
\centerline{\psfig{figure=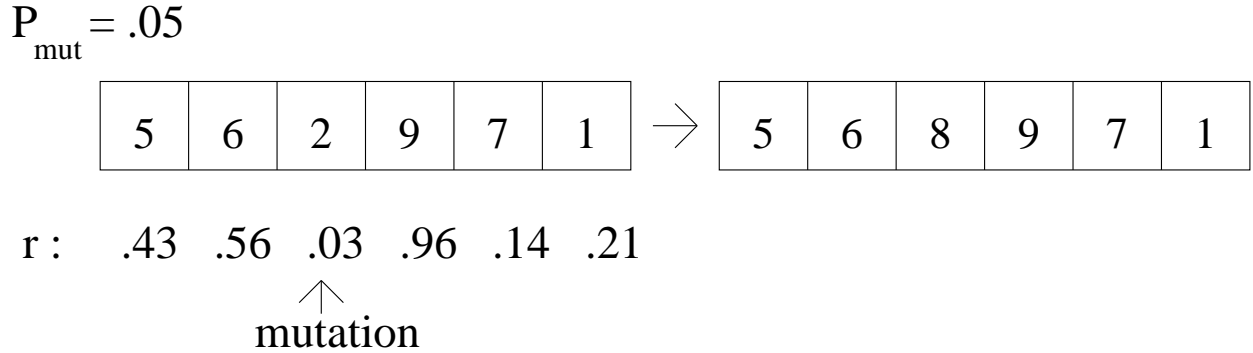,height=6.5cm}}
\caption{The mutation procedure. For each of the six locations along the string,
a random number $r$ between 0 and 1 is generated and compared with 
$p_{\rm mut}$. If $r$ is smaller
than $p_{\rm mut}$, a new, random value is assigned to the gene.}
\label{mutfig}
\end{figure*}

Thus, in this way, a new set of chromosomes is formed.
Usually not all new chromosomes are formed by crossover: Instead, some 
chromosomes, e.g. the ones with highest fitness, are copied directly as they are, and 
the rest are formed by crossover. 

Finally, mutation is applied to the new chromosomes. 
In order to perform mutation  on a chromosome, a random number $r$ is generated for 
each gene along the string, and the condition $r < p_{\rm mut}$, where $p_{\rm mut}$ 
is the mutation probability, is tested. If the condition is satisfied, the value 
of the gene is changed to a new random value. The procedure is illustrated in
Fig. \ref{mutfig}.

The chromosomes thus obtained (or, more strictly, the individuals corresponding to the
chromosomes) constitute the second generation. The individuals of the second generation
are then evaluated and fitness values are assigned to each individual, after which
the third generation is formed etc. This process continues until an acceptable solution 
to the problem has been found.

The description above only scratches the surface of the vast subject of GAs and the 
interested reader is again referred to the references cited at the beginning of the
Appendix. 

\begin{acknowledgements}
The author would like to thank Dr. K. J. Donner for carefully reading the
manuscript.
\end{acknowledgements}


\begin{thebibliography} {}
\bibitem[1995]{char} Charbonneau, P., 1995, ApJS, 101, 309
\bibitem[1988]{danb} Danby, J.M.A., 1988, Fundamentals of celestial mechanics,
William-Bell, Richmond
\bibitem[1991]{davi} Davis, L. (Ed.), 1991, Handbook of genetic algorithms,
Van Nostrand Reinhold, New York
\bibitem[1963] {deut} Deutsch, R., 1963, Orbital dynamics of space vehicles,
Prentice-Hall, Englewood Cliffs
\bibitem[1991] {enat} Engstr\"om, S., Athanassoula, E., 1991. In: Sundelius, B. (ed.)
Dynamics of disc galaxies, G\"oteborg University and Chalmers
University of Technology, G\"oteborg, p. 215
\bibitem[1996]{gibs} Gibson, S., Charbonneau, P., 1996, BAAS, 188, \# 36.22
\bibitem[1995]{haka} Hakala, P.J., 1995, A\&A, 296, 164
\bibitem[1990] {hern} Hernquist, L., 1990. In: Wielen, R. (ed.),
Dynamics and interactions of galaxies, Springer, Berlin, p. 108
\bibitem[1975]{holl} Holland, J.H., 1975, Adaptation in natural and artificial systems,
1st ed.: University of Michigan Press, Ann Arbor; 2nd ed.: 1992, MIT Press, Cambridge 
\bibitem[1990] {hoby} Howard, S., Byrd, G.G., 1990, AJ, 99, 1798
\bibitem[1995] {lang} Lang, M.J., Ir. Astron. J., 22, 167
\bibitem[1997] {lazi} Lazio, T.J., 1997, PASP, 109, 1068
\bibitem[1996]{mitc} Mitchell, M., 1996, An introduction to genetic algorithms, 
MIT Press, Cambridge 
\bibitem[1993]{sala} Salo, H., Laurikainen, E., 1993, ApJ, 410, 586
\bibitem[1995]{tcst} Tomczyk, S., Charbonneau, P., Schou, J., Thompson, M.J., 1995. In:
Hoeksema, J.T., Domingo, V., Fleck, B., Battrick, B. (eds.) 
Proc 4th SOHO Workshop: Helioseismology, ESA, Noordwijk, p. 271
\end{thebibliography}
\end{document}